\documentclass[10pt,letterpaper]{article}
\usepackage[top=0.85in,left=2.75in,footskip=0.75in]{geometry}

\usepackage{amsmath,amssymb}

\usepackage{changepage}

\usepackage{textcomp,marvosym}

\usepackage{cite}

\usepackage{nameref,hyperref}

\usepackage[right]{lineno}

\usepackage[nopatch=eqnum]{microtype}
\DisableLigatures[f]{encoding = *, family = * }

\usepackage[table]{xcolor}

\usepackage{array}

\newcolumntype{+}{!{\vrule width 2pt}}

\newlength\savedwidth

\raggedright
\usepackage[aboveskip=1pt,labelfont=bf,labelsep=period,justification=raggedright,singlelinecheck=off]{caption}

\makeatletter
\renewcommand{\@biblabel}[1]{\quad#1.}
\makeatother

\usepackage{lastpage,fancyhdr,graphicx}
\usepackage{epstopdf}
\fancyheadoffset[L]{2.25in}
\fancyfootoffset[L]{2.25in}
\usepackage{siunitx}
\usepackage{physics}

\usepackage{booktabs}
\usepackage{graphicx}

\usepackage{multirow}
\usepackage{indentfirst}
\usepackage{ragged2e}
\justifying
\usepackage{enumitem}

\begin{document}
\vspace*{0.2in}

\begin{flushleft}
{\Large
\textbf\newline{Physical model of serum supplemented medium flow in organ-on-a-chip systems} % Please use "sentence case" for title and headings (capitalize only the first word in a title (or heading), the first word in a subtitle (or subheading), and any proper nouns).
}
\newline
\\
Viesturs Šints\textsuperscript{1}*,
Jānis Cīmurs\textsuperscript{1},
Mihails Birjukovs\textsuperscript{1},
Ivars Driķis\textsuperscript{1},
Karīna Goluba\textsuperscript{2},
Kaspars Jēkabsons\textsuperscript{2},
Vadims Parfejevs\textsuperscript{2},
Una Riekstiņa\textsuperscript{2},
Gatis Mozoļevskis\textsuperscript{3},
Roberts Rimša\textsuperscript{3},
Guntars Kitenbergs\textsuperscript{1}
\\
\bigskip
\textbf{1} Laboratory of Magnetic Soft Materials, University of Latvia, Jelgavas street 3, Riga, Latvia
\\
\textbf{2} Faculty of Medicine and Life sciences, University of Latvia, Jelgavas street 3, Riga, Latvia
\\
\textbf{3} Micro and Nanodevices Laboratory, Institute of Solid State Physics, University of Latvia, Kengaraga street 8, Riga, Latvia
\\
\bigskip

* viesturs.sints@lu.lv

\end{flushleft}
\section*{Abstract}

Creating a physiologically relevant shear stress in organ-on-a-chip (OOC) devices requires careful tailoring of microfluidic flow parameters. It is currently fairly common to use a simple approximation assuming a constant viscosity, even for serum-based media. Here, we show that a popular nutrient solution (Dulbecco’s Modified Eagle Medium supplemented with Fetal Bovine Serum) requires a more complex treatment (i.e., is a non-Newtonian fluid), with observed shear stress values significantly greater than reported in literature. We measure the rheology of the solutions and combine it with a 3-dimensional flow field measurement to derive shear stress at the channel surface. We verify the experiments with numerical simulations, finding good agreement and deriving flow properties. Finally, we provide relevant expressions for shear stress approximation, suitable for development of OOC devices with various geometries.

% Please keep the Author Summary between 150 and 200 words
% Use first person. PLOS ONE authors please skip this step. 
% Author Summary not valid for PLOS ONE submissions.   
%\section*{Author summary}

%\linenumbers

% Use "Eq" instead of "Equation" for equation citations.

\section*{Introduction}

Organ-on-a-chip (OOC) systems are designed to mimic organ functionality in microfluidic devices \cite{bhatia_microfluidic_2014}. Operation of such devices often involves fluid flow and interaction of this flow with organ cells within the device. It is therefore important to understand the fluid properties. A crucial consideration is whether the flow medium in the OOC device is Newtonian (viscosity given by a single parameter regardless of the fluid shear rate) or non-Newtonian (fluid viscosity depends on the shear rate). Shear rate is a measure of the rate of change of velocity over the cross-section of a flow channel, with shear rate distribution in non-Newtonian fluids deviating significantly from that of Newtonian fluids.

Shear stress within a flow channel of a microfluidic device depends on both shear rate and viscosity, and is therefore likely to significantly vary if fluid viscosity depends on the shear rate. At the same time, shear stress is also associated with essential physiological effects \cite{espina_response_2023}, making it fundamental to multiple current and potential OOC applications. For example, in the case of blood vessels, shear stress can affect vascular morphogenesis, maturation and vessel permeability \cite{komarova, roux}. Endothelial cells reorganize their cytoskeleton and adjust the composition of intercellular junction proteins according to the shear stress experienced \cite{garcia-polite,tzima}. Similarly, in epithelial cells, certain fluid shear stress intensity can cause a variety of phenotypical, metabolic and functional alterations, including changes in epithelial layer permeability and cell maturation \cite{delon,lindner}. Moreover, forces due to shear stress are also involved in pathophysiology of some disorders, such as obstructive pancreatitis: prolonged stimulation of a mechanically responsive ion channel PIEZO-1 by high shear stress is sufficient to induce changes in pancreatic stellate cells and cause organ fibrosis \cite{swain2022}, while similar activation of acinar cells triggers pancreatitis \cite{swain2020}. Therefore, shear stress plays a crucial role in OOC system design.

The level of shear stress should be adjusted based on the type of cells it will affect. For cells that naturally experience shear stress, it is important to apply an optimal range to support their function. However, for cells that are sensitive to shear stress, it should be minimized to avoid detrimental effects. \cite{leung_guide_2022}. With this in mind, flow shear stress determination in OOC devices is clearly an important task. Relevant methods include using online tools offered by microfluidics companies \cite{calculator}, numerical simulations of fluid flow, and employing a relation between shear stress and flow rate in a channel of a known geometry (e.g.,  \cite{fois_dynamic_2021,shao_integrated_2009,kim_dynamic_2024}). Using either of the methods requires knowledge of the medium viscosity, which means one must classify the fluid as Newtonian or non-Newtonian. Note that relating the flow rate and shear stress via the Navier-Stokes equations, in the form used in the OOC research cited above, is only appropriate if the fluid is Newtonian.

To verify the fluid flow model assumed in literature, we perform our experiments with a popular commercial culture media: Dulbecco’s Modified Eagle Medium (DMEM), supplemented with Fetal Bovine Serum (FBS). The viscosity of DMEM supplemented with various concentrations of FBS, depending on the shear rate, has been measured before \cite{Poon}. However, the author of the paper does warn about possible non-Newtonian behavior for conditions not considered in the study (which limits itself to a range of shear rate values that may entirely not cover those encountered in OOC applications) and recommends repeated measurements at lower flow rates.

As the experimental measurement results presented in this paper will demonstrate, this nutrient solution is indeed non-Newtonian, and exhibits a shear thinning behavior. We will show that a power-law model can be used to describe the viscosity of the medium. The flow velocity and shear rate values presented here should be relevant to a variety of applications, including intestine and kidney cell OOCs \cite{corral-najera_polymeric_2023}.

Further, we will present measurements of flow fields within a channel of an OOC device, focusing on the non-Newtonian viscosity effects on the flow velocity distribution across the channel cross-section. We will introduce a mathematical description for the flow and present numerical modeling results, compare them to the experimental data. Finally, we will use the experimentally determined flow velocity and fluid viscosity to derive shear stress values at channel walls, allowing us to demonstrate the magnitude of the effect using the measured viscosity values, instead of relying on the data reported in literature under the assumption that DMEM supplemented by FBS is a Newtonian fluid.

Note that the Results section contains only a brief description of experimental, mathematical and numerical methods, for context. Details are provided in the Methods and Materials section.

\section*{Results}

\subsection*{Medium \& its viscosity}

\begin{figure}[ht]
    \centering
    \includegraphics[width=0.85\columnwidth]{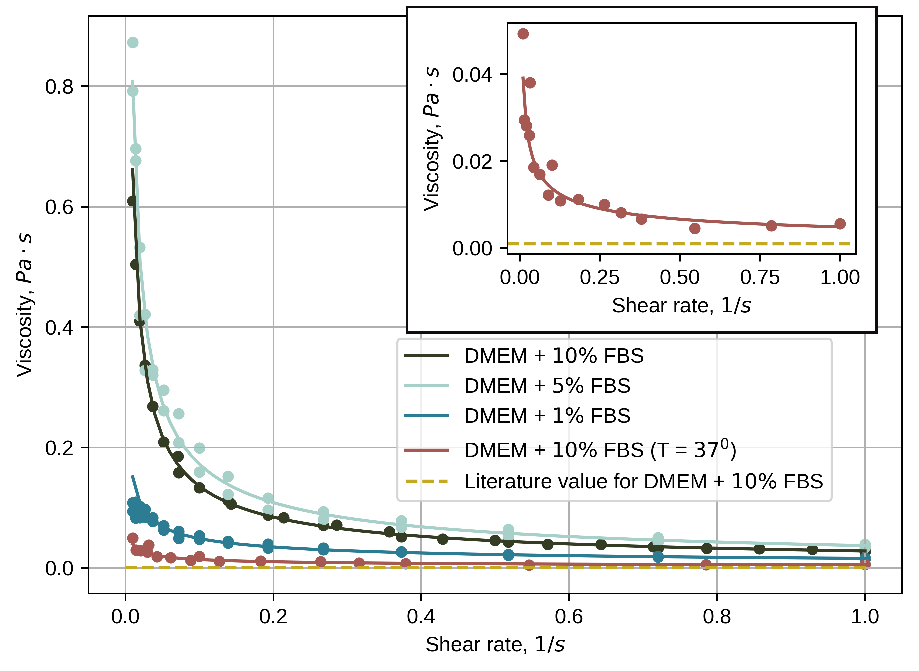}
    \caption{Viscosity of the medium over the shear rate range relevant to cell growth. The solid lines represent fits of experimental data to power law \eqref{eq:Ostwald}. Inset: viscosity of DMEM + $10 ~ \%$ FBS at $T=\SI{37}{\degreeCelsius}$ versus shear rate.}
    \label{fig:viscosity}
\end{figure}

The substance used for experimental investigation was DMEM supplemented with a variable amount (1\%, 5\% or 10\%) FBS and 1\% penicillin/streptomycin (all supplied by \textit{Gibco}). The viscosity of the medium was measured with a cone-plate rheometer, at room temperature ($T=\SI{20}{\degreeCelsius}$) and at $T=\SI{37}{\degreeCelsius}$. The viscosity dependence on the shear rate is shown in Fig. \ref{fig:viscosity}. The majority of the results are from the measurements at room temperature, with the inset focusing on a measurement of DMEM + $10 ~ \%$ FBS at $T=\SI{37}{\degreeCelsius}$. The latter dataset is the most relevant to actual cell cultivation. The dashed line in Fig. \ref{fig:viscosity} corresponds to the DMEM + $10 ~ \%$ FBS dynamic viscosity value provided in  \cite{Poon}, $\mu = \SI{9.3e-4}{\Pa \s}$. It should be noted that similar values are used in other papers as well, e.g., \cite{delon}, and therefore this baseline will be referred to as the \textit{literature value}.

One immediate observation is that, while the measured values do converge to the literature value as the shear rate increases, the viscosity at lower shear rates is much grater than the literature value. Note that the shear rate range shown in Fig. \ref{fig:viscosity} contains values that we would encounter in the OOC device used in our experiments -- please see Supporting information (\nameref{S1_Fig_shear_rate}) for the shear rate data.

Clearly, none of the samples are Newtonian fluids. Instead, the viscosity decreases with shear rate, which is known as shear thinning. For non-Newtonian fluids, the relation between shear rate and viscosity is given by the Ostwald-de Waele power law \eqref{eq:Ostwald}. The fit parameters for \eqref{eq:Ostwald} are summarized in Table \ref{tab:Ostwald}.

One can also note two trends: viscosity values increase with the FBS concentration, and increasing temperature decreases the viscosity, as expected. As seen in the inset of Fig. \ref{fig:viscosity}, even at the body temperature, viscosity exceeds the literature value over the entire shear rate range.

\subsection*{Fluid flow in the OOC device}

\begin{figure}[ht]
    \centering
    \includegraphics[width=0.85\linewidth]{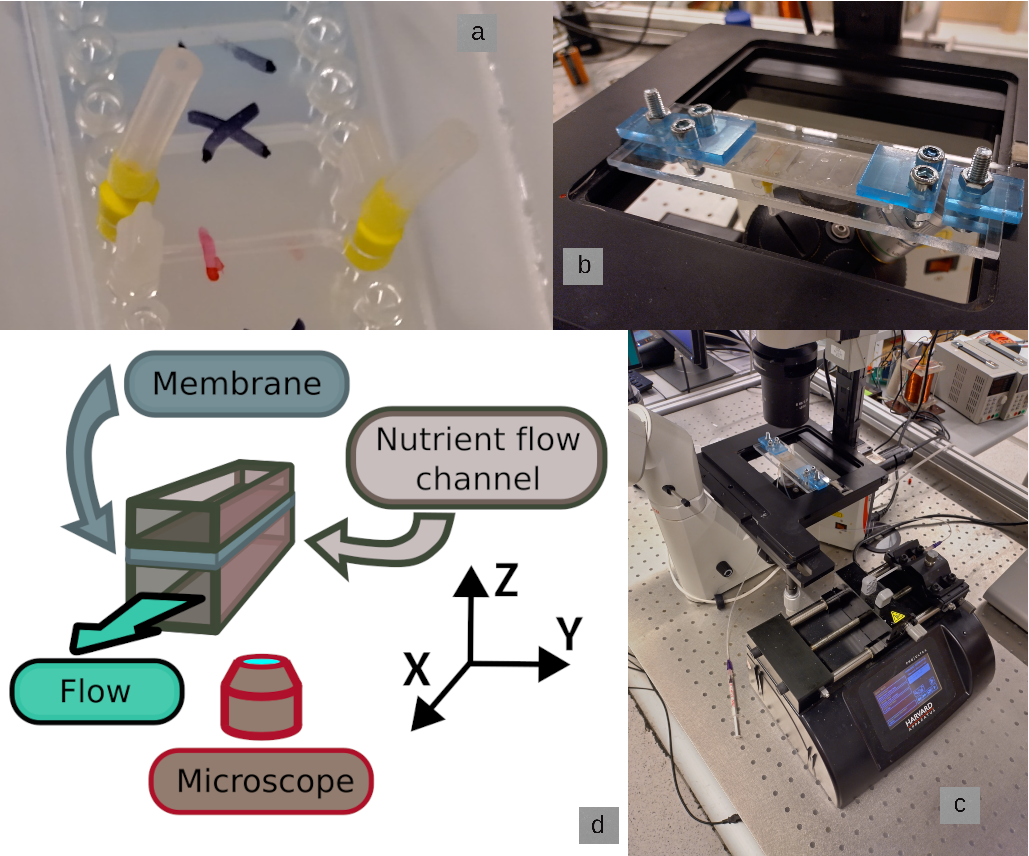}
    \caption{The experimental setup: (a) an array of channel pairs, the pair marked red has the inlet and outlet for one channel connected (the thicker, nutrient flow channel), and blocked for the other channel; (b) OOC device placed in a microscope; (c) the microscope with an OOC device and the syringe pump; (d) a schematic of the channel layout.}
    \label{fig:schem}
\end{figure}

We used OOC chips with pairs of vertically stacked channels separated by a PET membrane. We examine flow in one of the channels, $\SI{1.25}{\mm}$ high and $\SI{1.2}{\mm}$ wide, while the other, thinner channel, is filled with the same fluid, but its inlet and outlet are blocked. The flow is generated by a syringe pump, with flow rate set to $4$ or $\SI{8}{\ul\per\min}$. The flow rates correspond to the cell cultivation process reported with chips of similar geometry \cite{strods_development_2024}. The chips are held in a microscope using a fixture. A schematic of the resulting geometry is shown in Fig. \ref{fig:schem}, where coordinate conventions are also introduced.

\begin{figure}[ht]
    \centering
    \includegraphics[width=1.00\columnwidth]{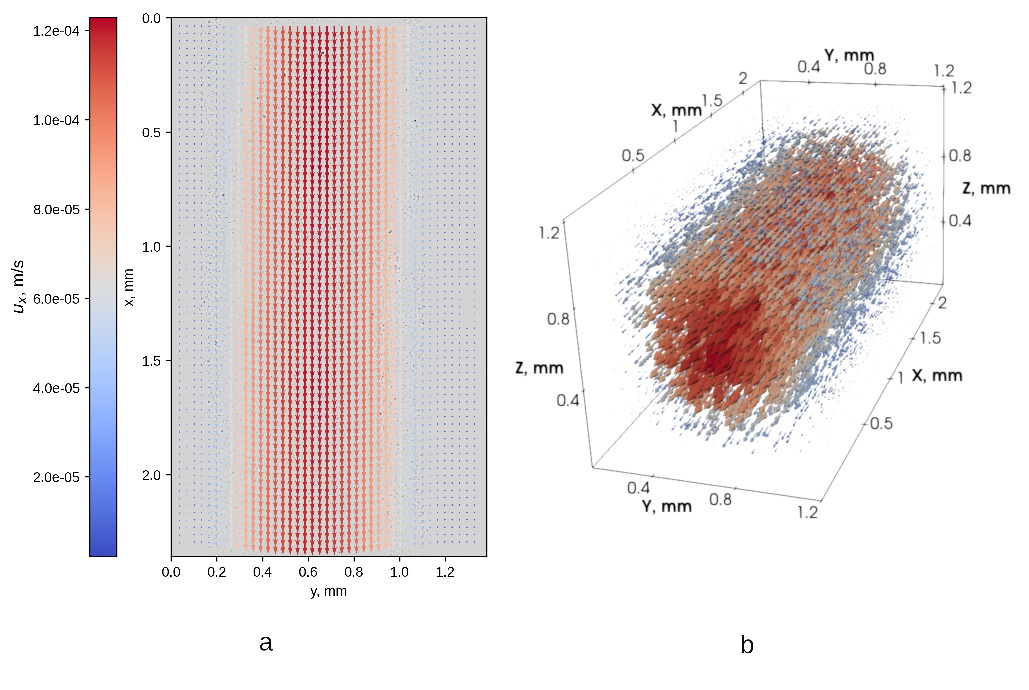}
    \caption{An example of a measured velocity field in the microchannel. Left: velocity distribution at a given $z$ coordinate value ($z=0.5 ~ mm$). Right: a vector field $\vec{u}(x,y,z)$ assembled from a stack of plane measurements for the channel volume within the microscope field of view.}
    \label{fig:flow3d}
\end{figure}

The flow field within the channels was measured using particle image velocimetry (PIV) at the microscope focus distance, giving us velocity distribution in $(y,x)$ coordinates, at a single $z$ coordinate, as shown in Fig. \ref{fig:flow3d} (a). We do this for every relevant $z$. Stacking these vector fields then gives us a flow field for the entire channel, as shown in Fig. \ref{fig:flow3d} (b).

\begin{figure}
    \centering
    \includegraphics[width=1\linewidth]{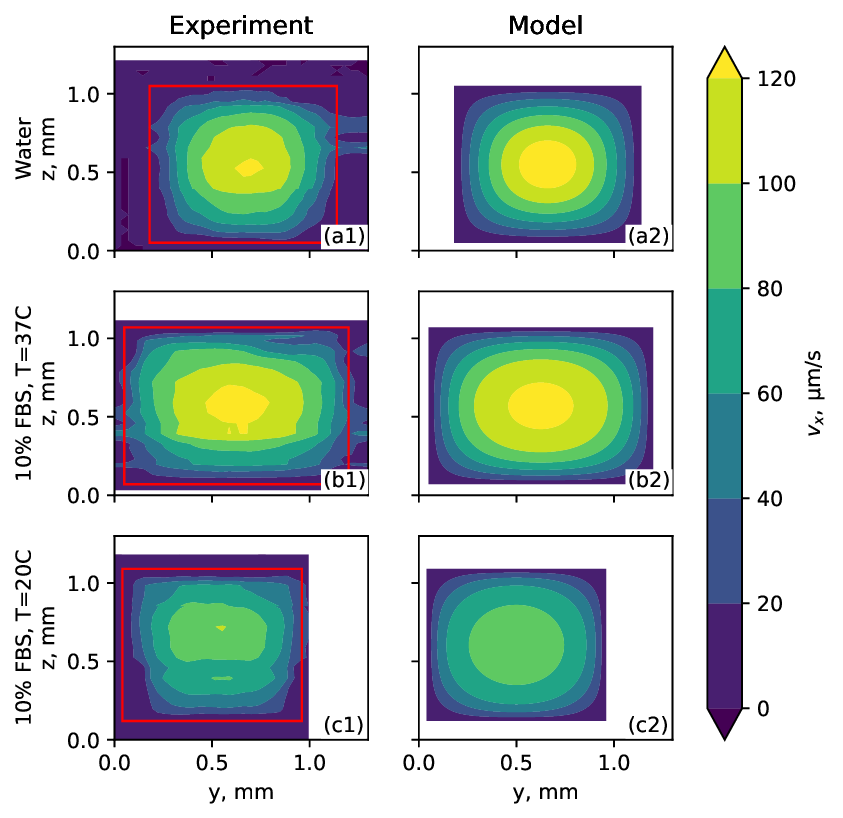}
    \caption{(a1), (b1), and (c1) -- Flow velocity obtained using PIV; (a2), (b2) and (c2) -- numerically calculated flow velocity using COMSOL. (a1) and (a2) -- water ($n=1$); (b1) and (b2) -- DMEM + 10\% FBS at $\SI{37}{\degreeCelsius}$ ($n=0.5$); (c1) and (c2) -- DMEM + 10\% FBS at $\SI{20}{\degreeCelsius}$ ($n=0.32$). Channel boundaries are indicated with red lines (zero velocity boundary condition). Here $n$ is the flow index in the power law \eqref{eq:Ostwald}.}
    \label{fig:2D_flow}
\end{figure}

Flow field measurements were performed for DMEM supplemented with all FBS concentrations considered in Fig. \ref{fig:viscosity}, as well as water, for reference. Cross-sections of the average flow fields in $(y,z)$ coordinates are shown in Fig. \ref{fig:2D_flow} (a1), (b1), and (c1). The corresponding numerical modeling results are shown in Fig. \ref{fig:2D_flow} (a2), (b2), and (c2), where the experimentally obtained dependence of fluid viscosity on shear rate has been employed. The samples represented here are water (Newtonian), DMEM supplemented by $10\%$ FBS as a more extreme example of a non-Newtonian fluid, and DMEM supplemented by $10\%$ FBS at $\SI{37}{\degreeCelsius}$. The latter is the most representative of a medium that would be used in OOC.

\begin{figure}
\begin{adjustwidth}{-2.25in}{0in}
    \centering
    \includegraphics[width=\linewidth]{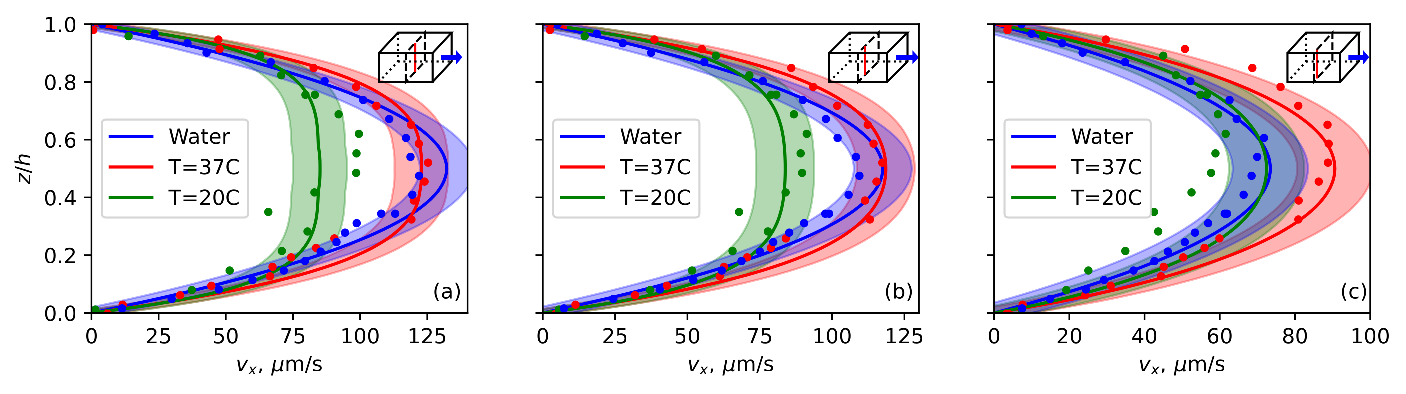}
    \caption{Velocity profiles obtained via PIV (dots) and COMSOL (solid lines) for 3 different fluids: water (blue, $n=1$), DMEM + 10\% FBS at $T=\SI{37}{\degreeCelsius}$ (red, $n=0.5$), DMEM + 10\% FBS at room temperature $T=\SI{20}{\degreeCelsius}$ (green, $n=0.32$). The profiles are compared in the middle of the channel (a), 1/3 of the channel width (b), and 1/6 of the channel width (c). Shaded regions are 10\% error margins for COMSOL. $n$ is the flow index in the power law equation \eqref{eq:Ostwald}.}
    \label{fig:1D_flow}
\end{adjustwidth}
\end{figure}

As seen in Fig. \ref{fig:2D_flow} and Fig. \ref{fig:1D_flow}, the non-Newtonian behavior of FBS-supplemented DMEM results in a flatter velocity profile at the center of the channel, with a less pronounced velocity peak at the center of the channel. This also means that the shear rate distribution in the channel is different from the case of a Newtonian fluid. This trend is observed in both experimental and numerical results, and the velocity maxima are in a reasonable agreement. Sources of measurement errors and discrepancies are discussed in the Methods and materials section. The red lines in Fig. \ref{fig:2D_flow} represent boundaries of the channel, identified as described in the Methods and materials section. The correspondence between theoretical and experimental values is further analyzed in Fig. \ref{fig:1D_flow}, where flow velocity profiles at several $y$ values are compared. The experimentally obtained velocity profiles agree with the theoretical model qualitatively and, within the error margins, also quantitatively.

\subsection*{Shear stress due to fluid flow}

We have by now demonstrated that DMEM supplemented by various amounts of FBS is a non-Newtonian fluid, with a viscosity that is a function of the fluid shear rate, and is significantly larger at low shear rate values than previously assumed in the literature. We have also obtained flow velocity fields from within the flow channels in an OOC device, and demonstrated the effect of the non-Newtonian nature on the velocity distribution. Flow shear stress, being a product of a shear rate and fluid viscosity, would be expected to be affected dramatically, compared to a Newtonian fluid with a viscosity like that reported for DMEM supplemented by FBS at higher shear rates (or assuming independence of shear rate).

\begin{figure}
    \centering
    \includegraphics[width=0.97\linewidth]{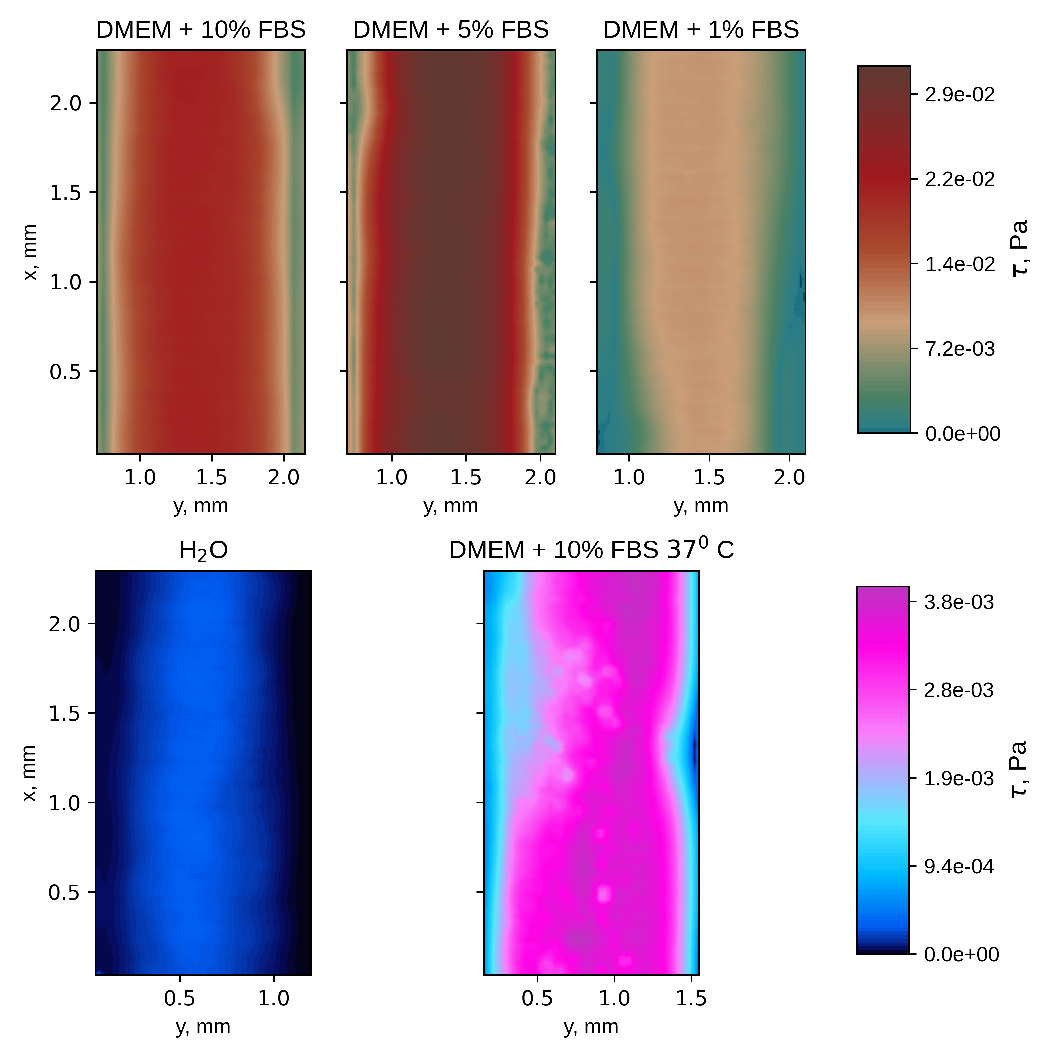}
    \caption{Shear stress at the membrane, which is the channel surface relevant for cell growth. Note the different color schemes in top and bottom rows, necessitated by the significant spread of of shear stress values.}
    \label{fig:shear_2d}
\end{figure}

The shear stress field is derived from the experimental results  by using the obtained velocity field and the shear rate to shear stress relations. We focus on the values near the membrane (the top channel wall) where cell growth would take place. The experimental shear stress fields near the membrane are shown in Fig. \ref{fig:shear_2d}. These fields exhibit a wide range of shear stress values -- note the different scales used for the top and bottom rows of Fig. \ref{fig:shear_2d}. The results most relevant for the OOC research are for DMEM + 10 $\%$ FBS at $\SI{37}{\degreeCelsius}$ temperature (corresponding to the OOC working conditions) and a Newtonian fluid, water.

For reference, we also compare the experimental values to the results we would obtain if we were to assume the nutrient solution to be a Newtonian fluid. For this, we introduce the concept of \textit{Newtonian DMEM + $10\%$ FBS}. As we have shown here, this is not a physically valid model of the fluid -- it is, however, currently the most common treatment in literature, hence we refer to it as the \textit{literature model}. To obtain shear stress for this case, we note that all Newtonian fluids exhibit identical velocity distributions under the same boundary conditions in the channel. If DMEM + $10\%$ FBS was Newtonian, it would have the same flow velocity distribution as water, so we use the velocity field measured for water as the result for a general Newtonian fluid. Viscosity is assumed to be constant and equal to $\mu = \SI{9.3e-4}{\Pa \s}$. One way to describe this approach is -- this is a description of DMEM + $10\%$ FBS that we would arrive at by following the literature data. The maximum shear stress on the membrane given by this model is compared to the results obtained experimentally and numerical modeling in Table \ref{tab:max_shear_stress}, showing an order of magnitude difference between the literature model and the actual DMEM + $10\%$ FBS values at $\SI{37}{\degreeCelsius}$.

\begin{table}[h]
    \centering
    \caption{Maximum shear stress values at the flow channel membrane where cell growth would occur: results from experimental data and numerical modeling. Note that the experimental values could be expected to have as much as $25\%$ measurement error}
    \begin{tabular}{lccc}
       \toprule
        \multirow{2}{*}{Sample} & $\tau_\text{max}$, \si{dyne\per\cm\squared} & \multicolumn{2}{c}{$\tau_\text{max}$, Pa} \\
         & Experimental & Experimental & COMSOL \\
        \midrule
        DMEM + $10\%$ FBS & 0.21 & 0.021 &  0.026\\
        DMEM + $5\%$ FBS & 0.31 & 0.031 & 0.033\\
        DMEM + $1\%$ FBS & 0.098 & 0.0098 & 0.013\\
        H$_2$O & 0.0052 & 0.00052 &  0.00052\\
        DMEM + $10\%$ FBS $\SI{37}{\degreeCelsius}$ & 0.046 & 0.0046 & 0.0039\\
        Literature & 0.0054 & 0.00054 & 0.00054\\
        \bottomrule
    \end{tabular}
    \label{tab:max_shear_stress}
\end{table}

From theoretical and numerical modeling results, we see that in a non-Newtonian fluid, shear stress $\tau$ exerted by the fluid is a power law function of the of the flow rate $Q$: \[\tau=K_\tau Q^{\frac{1}{n}}\] 
where $n$ is the flow index in \eqref{eq:Ostwald} and $K_\tau$ is a parameter depending on the channel and fluid. In a Newtonian fluid ($n=1$), shear stress is proportional to the flow rate. This and other scaling laws found in the Numerical simulations section allow comparing different experimental conditions.

As seen from simulation results in Figs. \ref{fig:2D_flow} and \ref{fig:1D_flow}, the velocity profile in a non-Newtonian fluid has a larger plateau in the middle of the channel. This results in a wider region of higher shear stress at the channel wall, as seen in Fig. \ref{fig:shear_2d}. From simulation results for a square cross-section, we know that the region of the wall with shear-stress not less than 90\% of the maximal value is in the middle 32\% of wall area for the Newtonian fluid; 47\% for the DMEM + 10\% FBS at $\SI{37}{\degreeCelsius}$ (body temperature); 52\% for the DMEM + 10\% FBS at $\SI{20}{\degreeCelsius}$ (room temperature). The area of the high shear region can be increased not only by switching from Newtonian to a non-Newtonian fluid, but also by using a much wider rectangular channel.

\section*{Discussion}

We have shown that DMEM supplemented by FBS is a non-Newtonian fluid with viscosity values diverging significantly at lower shear rates from a value referred to in multiple literature sources -- this constant, shear rate-independent value is only appropriate at shear rates in excess of those relevant to the OOC systems. This contradicts a significant amount of published research. At the same time, this conclusion is in agreement with Ref. \cite{Poon}, and agrees qualitatively with the conclusions in Ref. \cite{wang_new_2023}.

Investigation of rheological properties of the nutrient solution has shown that viscosity of the solution increases with FBS concentration. This suggests that FBS determines the non-Newtonian behavior of the solution. However, our experimental data presented here do not yet allow concluding this definitively.

We have developed a method for measuring the flow velocity field directly in an OOC device, and have shown a correspondence between the experimental and theoretical velocity values. The measured velocity profiles are within expectations for a non-Newtonian fluid, and provided us with the data required for shear stress calculation (viscosity parameters and the velocity field).

The non-Newtonian effects observed in the nutrient solution flow and the corresponding increase of shear stress are among the most significant results of this article. Shear stress measured in DMEM supplemented by $10 ~ \%$ FBS at $\SI{37}{\degreeCelsius}$ exceeds both the values seen for water and calculated using the viscosity value commonly found in literature by an order of magnitude. Results suggest that FBS content in the solution could strongly influence its rheological properties, and particularly shear stress exerted on the channel walls. The considerable discrepancy itself is noteworthy, even if the actual values might not be repeated at different geometries and flow rates. All of the above constitutes a strong argument against the use of (in this context) oversimplified shear rate-independent viscosity values and Newtonian fluid models for serum solutions.

We suggest the following approximate formula to calculate the flow rate $Q$ (in $\si{\ul\per\min}$ units) necessary to obtain the shear stress $\tau$ ($\tau_\text{SI}$ in $\si{\Pa}$ units and $\tau_\text{CGS}$ in $\si{dyn\per\cm\squared}$ units) for a channel with hydraulic diameter $a$ (in units $\si{\mm}$):
\[
    Q=a^3\cdot\left(\frac{\tau_\text{CGS}}{\num{2e-2}}\right)^{1.85}=a^3\cdot\left(\frac{\tau_\text{SI}}{\num{2e-3}}\right)^{1.85}
\]
For a square cross-section, $a$ is the side length; for a circular cross-section, $a$ is the diameter; for a rectangular cross-section with the height $h$ larger than the width $w$, $a=\sqrt[3]{h^2\cdot w}$. With this formula, the error should be $<15 \%$.

\section*{Methods \& materials}

\subsection*{Theoretical model}

\begin{figure}[ht]
    \centering
    \includegraphics[width=0.9\linewidth]{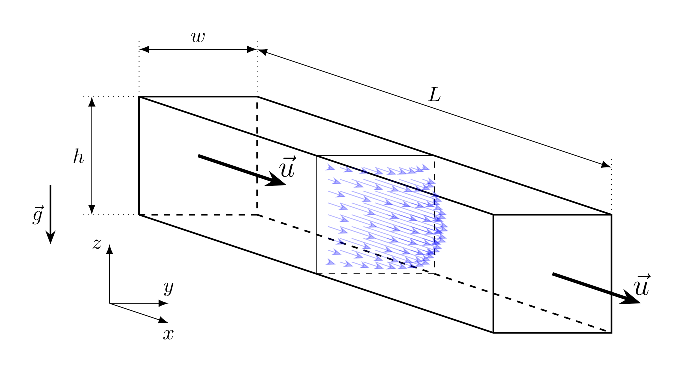}
    \caption{A schematic of the fluid flow in a rectangular channel with width $w$ ($y$-axis) and height $h$ ($z$-axis). The flow direction is $x>0$, and the direction of gravity $\vec g$ is $z<0$.}
    \label{fig:xyz}
\end{figure}

The fluid velocity field can be calculated using the Navier-Stokes equation. In microfluidics, channel dimensions are $\sim \ell=\SI{1}{\milli\meter}$, the flow rate is $<Q=\SI{10}{\ul\per\min}$, and kinematic viscosity is $\nu=\mu/\rho \gg \SI{1e-6}{\m\squared\per\s}$ (water), with the Reynolds number $Re \ll Q/\nu \ell=0.17$. This is known as creeping flow or Stokes flow, with negligible inertial effects. Therefore, here the fluid flow is described by the Stokes equation \cite{Stokes1851}, \cite[Chapter~4.8]{batchelor1967introduction}

\begin{gather}
    \nabla p=\nabla\cdot\left(\mu\nabla\vec u\right)\label{eq:Stokes}\\
    \nabla\cdot\vec u=0\text{ ,}\label{eq:continiuty}
\end{gather}
where $\vec u$ is the velocity field, $p$ is the pressure field, and $\mu$ is viscosity; $\nabla\vec u$ is the strain rate tensor, and $\mu\nabla\vec u$ is the shear stress tensor. The protein polymer chains can be extended by the flow in the $x$-direction (Fig. \ref{fig:xyz}), but that should not influence shear in the $yz$ plane. If the polymer chirality is not pronounced, clockwise and counterclockwise rotation should be equivalent. Therefore, we assume an isotropic fluid, with an isotropic relation between the shear rate and shear stress tensors, meaning $\mu$ is a scalar.

Due to the symmetries of the problem, stationary fluid flow in a channel with a constant cross-section can be expressed in Cartesian coordinates (please see axis definitions in Fig. \ref{fig:xyz}) using \eqref{eq:Stokes}:
\begin{equation}\label{eq:Stokes_xy}
    \pdv{y}\left(\mu\pdv{u_x}{y}\right)+
    \pdv{z}\left(\mu\pdv{u_x}{z}\right)=\dfrac{\delta p}{L}
\end{equation}

With the pressure gradient only along $x$-axis for a constant cross-section, \[u_z=u_y=0\] and from the continuity equation \eqref{eq:continiuty}, one has \[\pdv{u_x}{x}=0\] The pressure gradient is constant \[\pdv{p}{x}=\frac{\delta p}{L}<0\] and negative (since $u_x>0$), where $\delta p$ is the pressure difference between the ends of the channel.

The non-Newtonian behavior of a biological fluid containing many long protein chains can be described using the Ostwald–de Waele power law \cite{Ostwald1902}:
\begin{equation}\label{eq:Ostwald}
    \mu=K\dot{\gamma}^{n-1}
\end{equation}
where $K$ and $n$ are the flow consistency and the flow index, respectively, determined experimentally, and 
\begin{equation}
\label{eq:gamma_dot}
\dot\gamma=\sqrt{
\left(\pdv{u_x}{y}\right)^2+
\left(\pdv{u_x}{z}\right)^2
}
\end{equation}
is the second invariant of the strain rate tensor. Note that $n=1$ corresponds to a Newtonian fluid; if $0<n<1$, one has a shear-thinning non-Newtonian fluid, common for solutions with polymer chains; $n>1$ yields a shear-thickening non-Newtonian fluid, a behavior observed in granular matter. Here, we focus on a shear-thinning fluid with $0<n\leq 1$.

There exist more sophisticated models for stationary isotropic non-Newtonian fluid flow, e.g., the Carreau viscosity model \cite{Griffiths2020,carreau1972}, which works better for a larger strain rate range and does not have an infinite viscosity at zero strain rate. However, this and other more advanced models contain more parameters, which could result in overfitting the noisy experimental data instead of capturing the essential physics. Therefore, we use the power law relationship \eqref{eq:Ostwald}, which agrees reasonably well with the rheometer data.

\subsubsection*{Analytical solution in a circular channel}

To qualitatively understand the difference between non-Newtonian and Newtonian fluids, we show the simplest analytical solution, i.e., in a circular channel. In this setting, (\ref{eq:Stokes_xy}) simplifies to:
\begin{equation}
    \dfrac{1}{r}\dv{r}\left(r \mu\left(\dv{u_x}{r}\right)\cdot\dv{u_x}{r}\right)=\dfrac{\delta p}{L}\text{ ,}
\end{equation}
where $u_x=u_x(r)$ is a function of the radial distance $r$ from the symmetry axis of the circular channel; note that \[\mu=\mu\left(\dv{u_x}{r}\right)\] now depends only on the $u_x$ derivative with respect to $r$. Using boundary conditions $u_x(R)=0$, where $R$ is the radius of the circular channel, and \[\dv{u_x}{r}(r=0)=0\]the velocity distribution in the power law fluid \eqref{eq:Ostwald} can be obtained:
\begin{equation}
\label{eq:ux_circular}
  u_x(r)=\frac{n}{n+1}\left(\frac{\abs{\delta p}}{2KL}\right)^{\frac{1}{n}}\left(R^{\frac{1}{n}+1}-r^{\frac{1}{n}+1}\right)
\end{equation}

The corresponding flow rate is
\begin{equation}
\label{eq:Q_circular}
    Q=\int\limits_0^R u_x(r) \cdot 2\pi r \,\dd r =\frac{\pi R^3 \cdot n}{1+3n}\left(\frac{\abs{\delta p} R}{2KL}\right)^{\frac{1}{n}}
\end{equation}
and the shear-stress at the walls of the cylindrical channel produced by the power law fluid is uniform along the wall:
\begin{equation}
\label{eq:tau_circular}
    \tau=\frac{\abs{\delta p}R}{2L}=K\left(\frac{Q}{\pi R^3}\left(3+\frac{1}{n}\right)\right)^{n}
\end{equation}
where $n=1$ yields a Newtonian fluid with a Hagen–Poiseuille parabolic velocity distribution, with $K \rightarrow \mu$ \cite{stokes1845theories, hagen1839flow, poiseuille1844recherches}. One can see from (\ref{eq:tau_circular}) that shear-stress $\tau$ is linear in flow rate $Q$ only if the fluid is Newtonian ($n=1$). In a power law fluid, shear-stress $\tau$ scales non-linearly with $Q$.

\subsubsection*{Analytical solution in a rectangular channel with a small height}

In our experiments, the channel was rectangular, not cylindrical. Generally, a velocity field in a rectangular channel can not be obtained analytically. Therefore, numerical simulations are necessary. However, for a wide channel, an analytical solution exists. If the channel width is much larger than the height, the following velocity distribution between parallel plates can be assumed \cite{Griffiths2020}:
\begin{equation}
\label{eq:ux_rectangular}
    u_x(z)=\frac{n}{n+1}\left(\frac{\left|\delta p\right|}{KL}\right)^{\frac{1}{n}}\left(\left(\frac{h}{2}\right)^{\frac{1}{n}+1}-\left|\frac{h}{2}-z\right|^{\frac{1}{n}+1}\right)
\end{equation}

The corresponding flow rate is
\begin{equation}
\label{eq:Q_rectangular}
Q=w\int\limits_0^{h}u_x(z)\,\dd z=\frac{h^2 w \cdot n}{1+2n}\left(\frac{\left|\delta p\right| h}{2KL}\right)^{\frac{1}{n}}
\end{equation}

And the corresponding shear-stress on the upper and lower walls of the channel produced by the power law fluid is:
\begin{equation}\label{eq:tau_rectangular}
\tau=\frac{|\delta p| h}{L}=2K\left(\frac{Q}{h^2 w}\left(2+\frac{1}{n}\right)\right)^n
\end{equation}

\subsection*{Numerical simulations}

Fluid flow was simulated using a finite element method using the \textit{COMSOL Multiphysics}\textsuperscript{\textregistered} (COMSOL) software \cite{comsol}. The problem was solved for a 2D rectangular geometry using the coefficient form PDE. The solved equation is \eqref{eq:Stokes} with an isotropic coefficient given by \eqref{eq:Ostwald}. Dirichlet boundary conditions are set for all boundaries. Channel dimensions correspond to the experiments. The velocity field was initialized as follows 
\[u_x(y,z)=\frac{4u_{max}\cdot y z (w-y)(h-z)}{w^2 h^2}\] 
where $u_{max}$ is the maximal flow speed of the fluid flow from the experiment with similar flow rate. Extra fine physics-controlled mesh was used. The problem was solved using a stationary solver. The $K$ and $n$ values chosen according to table \ref{tab:Ostwald} and the pressure gradient $\frac{\delta p}{L}$ is chosen to match the flow rate with the experiments using the scaling laws presented below.

The results in a circular channel and a channel with a small height yield the following scaling laws:
\begin{itemize}[noitemsep]
    \item Shear stress $\tau$ is proportional to pressure gradient $\delta p/L$ and linear dimension of the channel $h$: $\tau =C_{s}\cdot h\cdot \delta p/L$, where $C_{s}$ is a channel shape-dependent dimensionless constant.
    \item The velocity profile $u_x(y,z)$ is proportional to the flow rate $Q$ and inversely proportional to the cross-section area $A$: $u_x(y,z)=Q/A\cdot f_{s}(y,z)$, where $f_s(y,z)$ is a shape-dependent dimensionless function.
    \item Shear stress $\tau$ is a non-linear function of flow rate $Q$ and the linear dimension of the channel $R$: $\tau= K\cdot C_{\tau}\cdot (Q/R^3)^{\frac{1}{n}}$, where $C_\tau$ is a shape-dependent dimensional parameter and values $K$ and $n$ are the viscosity parameters \eqref{eq:Ostwald}.
\end{itemize}

These scaling laws were tested using COMSOL for rectangular channels with different cross-section aspect ratios.

\subsection*{Fabrication \& geometry of the OOC device}

The chips have a vertically stacked design with the top channel height of $\SI{1.25}{\mm}$, designated for epithelial cells, and the bottom channel height of $\SI{0.20}{\mm}$, for endothelial cells. This distinction refers to the intended uses of both channels, whereas in the experiments described here, no cells are present in either of the channels. Both channels have an overlapping area of $\SI{18}{\mm\squared}$, with channel widths of $\SI{1.2}{\mm}$ and $\SI{1.0}{\mm}$, respectively. A schematic of the resulting geometry is shown in Fig. \ref{fig:schem}.

A syringe pump is used to generate flow in the larger of the channels, while the smaller channel is filled with the same fluid that is used for flow, with blocked inlets and outlets. The channels are separated by a $\SI{3}{\um}$ porous polyethylene terephthalate (PET) membrane ($\SI{3}{\um}$ pore size, $\SI{0.8e6}{pores\per\cm\squared}$ and $5.7 \%$ porosity, manufactured by \textit{it4IP}, Belgium).

The microfluidic chips were designed in \textit{Solidworks} (\textit{DS Solidworks Corp.}, USA) and fabricated using the off-stoichiometry thiol-ene (OSTE) and cyclo-olefin copolymer (COC). We used COC mini luer port microscope slides for the top channels and standard COC microscope slides for the bottom layers (both by \textit{Microfluidic ChipShop}, Germany). Master molds for the OOC channels were 3D printed using a masked stereolithography (MSLA) printer (\textit{Zortrax Inkspire}, \textit{Zortrax}, Poland) and ivory resin. The master mold was cleaned in an ultrasonic bath with isopropyl alcohol for 10 minutes, blow-dried with nitrogen, and then fully cured with UV light ($\SI{6.6}{\mW\per\cm\squared}$ for 30 minutes), followed by thermal treatment at $\SI{60}{\degreeCelsius}$ for 48 hours. PDMS was then mixed in at a 10:1 ratio, poured into the molds, and cured at $\SI{60}{\degreeCelsius}$ overnight.

For device fabrication, an OSTE 322 mixture (\textit{Mercene Labs}, Sweden) was prepared and processed in a \textit{Thinky} mixer at $\SI{750}{rpm}$ for 5 minutes in both mixing and defoaming modes. The mixture was then degassed in a vacuum desiccator for approximately 20 minutes. The surfaces of the COC slides were oxidized for 2 minutes using a plasma asher (\textit{GIGAbatch 360 M, PVA Tepla}, USA) at $\SI{600}{\W}$ with an oxygen flow rate of 800 sccm. Device assembly is finalized by injecting the OSTE mixture into the PDMS molds for each side, and exposing them to UV light using a \textit{Suss MA6} mask aligner ($\SI{850}{\mJ}$ bottom layer, $\SI{925}{\mJ}$ top layer). The COC slide with the exposed OSTE was then pressed against the membrane, and cured on a hot plate at $\SI{60}{\degreeCelsius}$ for 1 hour, with PTFE films and a $\SI{2}{\kg}$ weight pressing the devices onto the hot plate. After a similar preparation of the other layer, the two layers were aligned using an adapted alignment tool and pressed together to remove air pockets. The assembled OSTE/COC hybrid device was compressed from both sides and cured in an oven at $\SI{60}{\degreeCelsius}$ overnight.

\subsection*{Viscosity measurements}

Rheological properties of the fluids were measured with a \textit{Anton Paar MCR 502} rheometer, in a cone plate mode, with a constant shear rate across the sample. Peltier heating was used for temperature control, with measurements performed at room temperature ($T=\SI{20}{\degreeCelsius}$) and $T=\SI{37}{\degreeCelsius}$.

The mixtures of DMEM and all concentrations of FBS used here were found to exhibit a significant startup viscosity, with up to $\SI{2000}{\s}$ of probe rotation needed for the viscosity to become stationary for a measurement at a constant shear rate. An initialization procedure with $\SI{2000}{\s}$ of rotation at $\SI{1}{\s^{-1}}$ shear rate was performed prior to viscosity measurements. We assume this replicates the conditions in the substance flow within the tubing leading to the OOC device, and that it is therefore the stable plateau viscosity, rather than the transient startup value, that is relevant to the flow in the OOC device.

The measurements involved subjecting the sample fluid to rotation of the conical probe, with each measurement point corresponding to a $\SI{160}{\s}$ measurement of fluid viscosity at a given shear rate. Each sample was tested for a range of shear rates, focusing on measurements in the $\dot \gamma = 0.01 ~ - ~ \SI{1}{s^{-1}}$ range. The overall range of shear rates covered experimentally was $\dot \gamma = 0.01 ~ - ~ \SI{100}{s^{-1}}$. Measurements at shear rates below $\dot \gamma = \SI{0.01}{s^{-1}}$ were unreliable.

To estimate repeatability of the results, measurements were made with different variations of the samples, including different samples with the same FBS concentration, repeated measurements for the same sample, including a refill of the rheometer, and measuring the same sample over consecutive days. We provide an example in Supporting information (\nameref{S2_Fig_vicosity}). These measurements yield a $20 \%$ error estimate for the data obtained with parameters $K$ and $n$ in \eqref{eq:Ostwald}. All the values of viscosity used in experimental analysis are taken from the sample used in the respective experiment, with no more than $\SI{24}{\hour}$ having passed between the viscosity measurement and the flow experiment.

\begin{table}[ht]
\caption{Fit results for $K$ and $n$ values in \eqref{eq:Ostwald} for the experimental rheometry data shown in Fig. \ref{fig:viscosity}.}
\label{tab:Ostwald}
\begin{tabular}{llrr}
\toprule
 Sample & K & n \\
\midrule
DMEM+10\% FBS & 0.028 & 0.31 \\
DMEM+$5 \%$ FBS & 0.037 & 0.33 \\
DMEM+$1 \%$ FBS & 0.016 & 0.51 \\
DMEM+$10 \%$ FBS ($\SI{37}{\degreeCelsius}$) & 0.005 & 0.54 \\
\bottomrule
\end{tabular}
\end{table}

\subsection*{Particle image velocimetry \& image acquisition}

The PIV method is a type of image correlation velocimetry. The essence of the method involves taking two images of tracer particles within a fluid in quick succession, and measuring the displacement of these particles between the two images. This provides information about the velocity field of the fluid.

We use fluorescent $\SI{1.94}{\um}$ \textit{Nile Red} particles (\textit{Spherotech FH-2056-2}) mixed into the DMEM and FBS solution (or water), and subjected to ultrasound treatment. Viscosity measurements after tracer particle integration indicated that the solution properties were unaltered. We found that there was a risk of sedimentation for larger tracer particles, which limited maximum experiment runtime.

The measurements were performed using an inverted microscope (\textit{Leica DMI3000B}) with a $4\times$ magnification objective, and a PIV setup by \textit{Dantec Dynamics}, including the proprietary imaging software. A dual power 50-50 $2 \times \SI{50}{\mJ}$, $\SI{532}{\nm}$ laser was used to induce the tracer particle fluorescence prior to image acquisition. Images pairs were acquired in $\SI{0.06}{\ms}$ intervals with a $\SI{50}{\Hz}$ frequency.

For image acquisition, the microscope is focused at a point with a user-defined $z$ coordinate, following the definitions in Fig. \ref{fig:schem}. The first point is typically placed below the lower wall of the channel to make sure it is properly identified. Similarly, the last measurements are done at $z$ values above the channel and inside the membrane. A sequence of 100 image pairs is then acquired. This constitutes a measurement for one $z$ value, or one flow layer. Following that, the microscope focus is changed to a different $z$ value, and the next 100 image pair sequence is acquired, giving us the subsequent flow layer. The shift in microscope focus between two layers is $\delta z = \SI{33}{\um}$, although $\delta z = \SI{66}{\um}$ was frequently used for central areas of the channel. This typically resulted in 16-25 layers per full channel measurement. The images were taken near the middle of the channel $x$ axis, as far as possible from flow inlets and outlets.

The setup is enclosed in a heat insulating casing for temperature control. As in viscosity measurements, velocimetry is performed at room temperature and $T=\SI{37}{\degreeCelsius}$. Temperature inside the casing is regulated with a heater and a pair of \textit{DS18B20} temperature sensors with an \textit{Arduino} board. One of the sensors is placed near the chip, and the other between the chip and the syringe pump to ensure temperature uniformity within a $\delta T = \pm \SI{2}{\degreeCelsius}$ error.

\subsection*{Image processing}

Processing of the obtained images is crucial for applying the described PIV method for velocity field measurements. Image quality is affected by the suboptimal (from the point of view of PIV analysis) tracer particle choice. However, the primary challenge involves discriminating the observed particles by the layer of flow they are a part of. While adjustment of microscope focus brings one flow layer into focus, particles at different $z$ values can still contribute to the signal, either as background luminescence or discernible entities moving at the velocity of their respective layer. Therefore, pre-processing of the raw images to enhance the signal-to-noise and contrast-to-noise ratio (SNR/CNR, respectively) is required to obtain reliable results.

Two approaches are used for image processing. The first one involves a set of fairly standard procedures. Image gamma correction and histogram stretching are used to enhance the visibility of the tracer particles, followed by a successive application of low-pass and high-pass filters. All image transformations are performed in \textit{Python}. In the other approach, image processing is performed using \textit{Wolfram Mathematica} -- the code is open-source and is available on \textit{GitHub}: \href{https://github.com/Mihails-Birjukovs/OOC_particle_flow_CNR_boost_for_PIV/}{Mihails-Birjukovs/OOC\_particle\_flow\_CNR\_boost\_for\_PIV}. This approach is explained in more detail in the following chapters.

The raw images are $1344 \times \SI{1024}{px\squared}$ 16-bit grayscale TIFs, with $\sim 9$ pixels per particle (PPP). Their field of view contains the fluid flow channel and the surrounding background, so channel area segmentation is performed. For this, the minimum pixel-wise temporal projection for the image sequence is computed, and the resulting image is inverted. To enhance the channel wall CNR, soft color tone map masking (SCTMM) is applied \cite{birjukovs2021resolving, birjukovs-particle-EXIF, birjukovs-solidification-image-processing}, and then the initial channel wall mask is obtained via local adaptive binarization \cite{local-adaptive-thresholding}, with the local thresholding radius given by the median of image dimensions times a control parameter. To remove small-scale segmentation artifacts, a small-radius Gaussian filter is applied with subsequent re-binarization with the Otsu method \cite{otsu-thresholding}, followed by morphological closing with a disk kernel \cite{images-mathematical-morphology}. Finally, to eliminate any remaining artifacts at the channel boundary, morphological dilation (disk kernel) \cite{images-mathematical-morphology} is applied to slightly extend the boundary, and the non-border segments are removed from the final mask.

Images are cropped to the channel boundaries automatically, and image background correction is performed as follows: each image in a sequence is inverted, divided by the sequence mean image (a flat-field correction, FFC, with mean reference), inverted again, and then color tone mapping (CTM) is applied \cite{reproduction-of-color-chapter-6}. This correction method efficiently flattens the image luminance distribution and removes vignetting at the channel boundaries, as well as eliminates image artifacts due to reflections from the channel walls. Then the resulting images are multiplied by an inverted channel mask to isolate the particle flow region for PIV.

Corrected images typically still exhibit a rather low CNR due to haze and dynamic large-scale features moving with the particle flow (essentially, correlated noise). Therefore, dehazing is performed first via two iterations of the following: an input image is inverted, normalized, then CTM, normalization and reference-less FFC \cite{wolfram-brightness-equalize} are applied, followed by image inversion. Afterward, correlated noise and large-scale textures are removed using the non-local means masking (NMM) method \cite{birjukovs-particle-EXIF, birjukovs-solidification-image-processing}. Both operations significantly boost the particle CNR, but the SNR is slightly reduced, so Perona-Malik anisotropic diffusion \cite{perona-malik, weickert-nonlinear-aniso-diff-schemes} is applied to boost the image SNR while preserving particle CNR. Finally, to better resolve finer particles, CNR is boosted further by applying 15 iterations of the following: invert input images, apply CTM, invert the result. Optionally, one can obtain particle binary masks by applying the Kapur's segmentation algorithm \cite{kapur-entropy-segmentation}, followed by boundary segment removal and size thresholding. The resulting filtered images can be used as PIV input. Compared to the other approach, this more advanced method was found to provide a better distinction for particles belonging to a particlural flow layer, and sharper channel edge detection.

\subsection*{Measuring the velocity field}

PIV analysis is performed after image processing. The results shown here were obtained with the \textit{OpenPIV} software for \textit{Python}, although analysis using the \textit{PIVlab} plugin for \textit{MATLAB} was also performed for most of the relevant experimental data. Both methods yielded similar results. The PIV resolution (the smallest interrogation window size) was set to $\SI{32}{px}$, resulting in a $\delta y = \SI{37}{\um}$ velocity field resolution, roughly equal to the distance between the flow layer measurements ($\delta y \approx \delta z$).

The physical dimensions of the channel are determined either by the point where flow velocity drops to zero (the first and last layer to have non-zero velocities detected in them are assumed to be the layers near top and bottom walls, while side walls can be identified by velocity distribution by $y$ reaching zero) or by using the channel masks obtained by image processing (side walls only).

In attempting to estimate the precision of flow measurements, two key sources of error have been identified -- fluid flow fluctuations and an inconsistency of channel dimensions. The former is likely due to bubble formation, channel blockage, or some other defect along the flow path. As the flow fields corresponding to each $z$ are obtained in succession, such fluctuations would affect one or several layers of measurement, effectively translating into spatial defects seen in data. In practice, this is the primary source of the jagged edges of the velocity contours seen in Fig. \ref{fig:2D_flow}. We have observed velocity deviation to reach 7\% of the maximum velocity value. Inconsistencies in channel geometry are another major source of discrepancy between theory and experiment. The theoretical/numerical model assumes the channel is straight, with a rectangular cross-section. In experiments, the channel walls are slightly jagged, and the channel cross-section is only approximately rectangular. Due to the channel width variations and given that the top and bottom walls are not visible, the positions of the walls and the area of the cross section of the channel will not be consistent for all $x$ values when averaging over the measured flow field. We estimate the error of the channel width at $\sim 5\%$ of the average width.

\section*{Supporting information}

% Include only the SI item label in the paragraph heading. Use the \nameref{label} command to cite SI items in the text.

\paragraph*{S1 Fig.}
\label{S1_Fig_shear_rate}
{\bf Shear-rate in a square channel.} Simulated data of the shear rate $\dot{\gamma}$ \eqref{eq:gamma_dot} in the middle of the channel ($y=\SI{0.5}{\mm}$) for a channel with a square cross-section of size $\SI{1}{\mm}\times\SI{1}{\mm}$. The flow rate in the channel is $Q=\SI{4}{\ul\per\min}$. It can be seen that shear rate does not exceed $\dot\gamma=\SI{1}{1\per\s}$ for all fluids discussed in this article: Newtonian fluid (Water, blue line) ($n=1$); non-Newtonian fluid with $n=0.54$ (DMEM+$10 \%$ FBS  at $\SI{37}{\degreeCelsius}$ and DMEM+$1 \%$ FBS  at room temperature, red line); non-Newtonian fluid with $n=0.31$ (DMEM+$10 \%$ FBS and DMEM+$5 \%$ FBS at room temperature, green line). $n$ values correspond to power-law exponent \eqref{eq:Ostwald}. The bump near the center, pronounced by the red line, is due to nonphysical infinite viscosity at zero shear rate of the Ostwald formula \eqref{eq:Ostwald}. This bump has inessential influence on the flow profile and shear-rate on the wall. Less pronounced bump is by the green line.

\begin{figure}[ht]
\centering
    \includegraphics[width=0.9\linewidth]{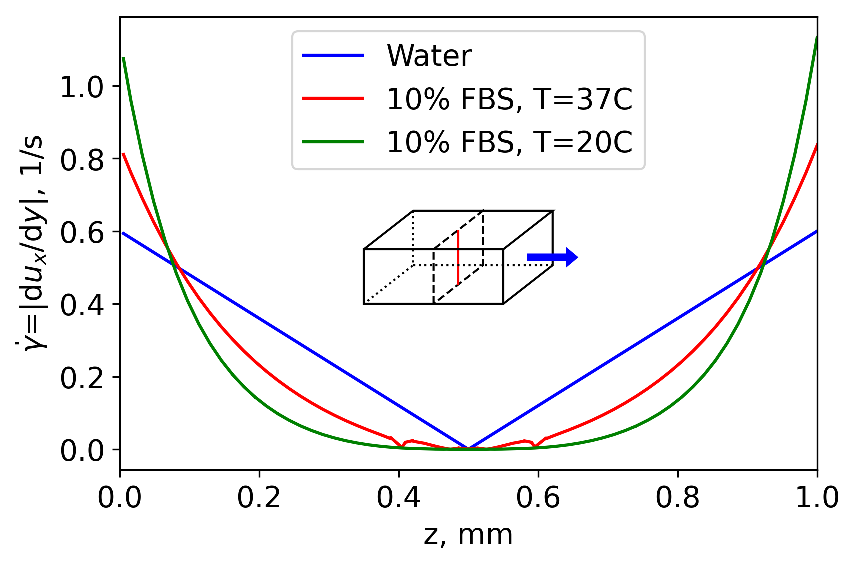}
\end{figure}

\paragraph*{S2 Fig.}
\label{S2_Fig_vicosity}
{\bf Viscosity measurement data.} Data from viscosity measurements for DMEM + 10 \% FBS. Here we have reviewed three samples, performing measurements multiple times (denoted by "runs"), including a refill of the same sample and measurements performed on different days. Ostwald-de Waele equation was fit to the entire ensemble of data points. A 20\% deviation from the fit line is included as a measure of repeatability of results.

\begin{figure}[ht]
    \centering
    \includegraphics[width=0.9\linewidth]{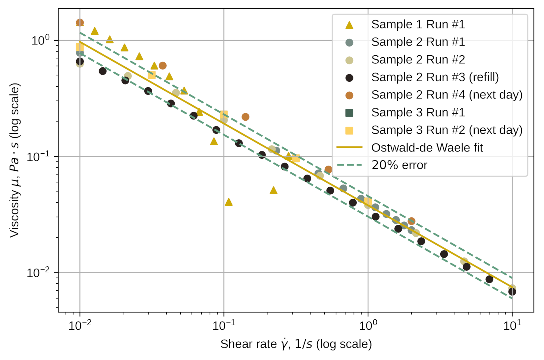}
\end{figure}

\paragraph*{S3 Fig.}
{\bf Comparison between flow rates.} Here we show the experimental velocity values of the fluid in the middle line of the channel for two different flow rates $Q=\SI{4}{\ul\per\min}$ and $Q=\SI{8}{\ul\per\min}$. We demonstrate here that increasing flow rate 2 times, the fluid velocity increases two times. The model predicts the same shape of the velocity distribution for both flow rates.

\begin{figure}[ht]
    \centering
    \includegraphics[width=0.7\linewidth]{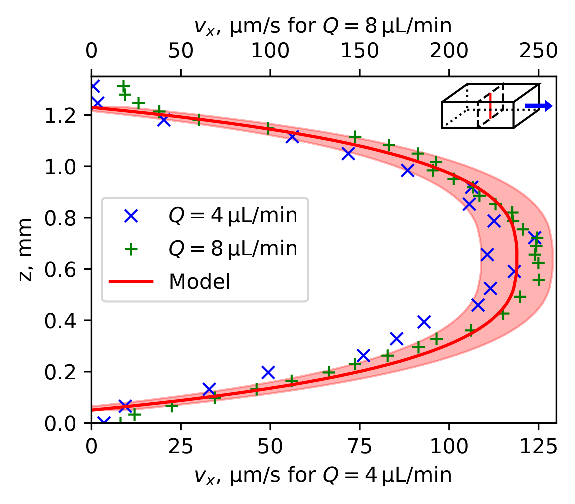}
\end{figure}

\section*{Acknowledgments}
This research is funded by the Latvian Council of Science State research program project “Smart Materials, Photonics, Technologies and Engineering Ecosystem”, No. VPP-EM-FOTONIKA-2022/1-0001. The authors also acknowledge the LLC "MikroTik" donation project no. 2259 "Simulations for stimulation of science", administered by the University of Latvia foundation, for providing the resources for numerical modeling.

\nolinenumbers

% Either type in your references using
% \begin{thebibliography}{}
% Text
% \end{thebibliography}
%
% or
%
% Compile your BiBTeX database using our plos2015.bst
% style file and paste the contents of your .bbl file
% here. See http://journals.plos.org/plosone/s/latex for 

\bibliography{bibliography}

\begin{thebibliography}{10}

\bibitem{bhatia_microfluidic_2014}
Bhatia SN, Ingber DE.
\newblock Microfluidic organs-on-chips.
\newblock Nat Biotechnol. 2014;32(8):760--772.
\newblock doi:{10.1038/nbt.2989}.

\bibitem{espina_response_2023}
Espina JA, Cordeiro MH, Milivojevic M, Pajić-Lijaković I, Barriga EH.
\newblock Response of cells and tissues to shear stress.
\newblock Journal of Cell Science. 2023;136(18):jcs260985.
\newblock doi:{10.1242/jcs.260985}.

\bibitem{komarova}
Komarova YA, Kruse K, Mehta D, Malik AB.
\newblock Protein interactions at endothelial junctions and signaling mechanisms regulating endothelial permeability.
\newblock Circ Res. 2017;120:179–206.
\newblock doi:{10.1161/CIRCRESAHA.116.306534}.

\bibitem{roux}
Roux E, Bougaran P, Dufourcq P, Couffinhal T.
\newblock Fluid Shear Stress Sensing by the Endothelial Layer.
\newblock Front Physiol. 2020;11.
\newblock doi:{10.3389/fphys.2020.00861}.

\bibitem{garcia-polite}
Garcia-Polite F, Martorell J, Del Rey-Puech P, Melgar-Lesmes P, O’Brien CC, Roquer J, et~al.
\newblock Pulsatility and high shear stress deteriorate barrier phenotype in brain microvascular endothelium.
\newblock J Cereb Blood Flow Metab. 2017;37:2614–2625.
\newblock doi:{10.1177/0271678X16672482}.

\bibitem{tzima}
Tzima E, Del~Pozo MA, Kiosses WB, Mohamed SA, Li S, Chien S, et~al.
\newblock Activation of Rac1 by shear stress in endothelial cells mediates both cytoskeletal reorganization and effects on gene expression.
\newblock EMBO J. 2002;21:6791–6800.
\newblock doi:{10.1093/emboj/cdf688}.

\bibitem{delon}
Delon LC, Guo Z, Oszmiana A, Chien CC, Gibson R, Prestidge C, et~al.
\newblock A systematic investigation of the effect of the fluid shear stress on Caco-2 cells towards the optimization of epithelial organ-on-chip models.
\newblock Biomaterials. 2019;225:119521.
\newblock doi:{10.1016/j.biomaterials.2019.119521}.

\bibitem{lindner}
Lindner M, Laporte A, Block S, Elomaa L, Weinhart M.
\newblock Physiological Shear Stress Enhances Differentiation, Mucus-Formation and Structural 3D Organization of Intestinal Epithelial Cells In Vitro.
\newblock Cells. 2021;10:2062.
\newblock doi:{10.3390/cells10082062}.

\bibitem{swain2022}
Swain SM, Romac JMJ, Vigna SR, Liddle RA.
\newblock Piezo1-mediated stellate cell activation causes pressure-induced pancreatic fibrosis in mice.
\newblock JCI Insight. 2022;7.
\newblock doi:{10.1172/jci.insight.158288}.

\bibitem{swain2020}
Swain SM, Romac JMJ, Shahid RA, Pandol SJ, Liedtke W, Vigna SR, et~al.
\newblock TRPV4 channel opening mediates pressure-induced pancreatitis initiated by Piezo1 activation.
\newblock J Clin Invest. 2020;130:2527–2541.
\newblock doi:{10.1172/JCI134111}.

\bibitem{leung_guide_2022}
Leung CM, De~Haan P, Ronaldson-Bouchard K, Kim GA, Ko J, Rho HS, et~al.
\newblock A guide to the organ-on-a-chip.
\newblock Nat Rev Methods Primers. 2022;2(1):33.
\newblock doi:{10.1038/s43586-022-00118-6}.

\bibitem{calculator}
FLUIGENT. Shear stress calculator; 2024.
\newblock Available from: \url{https://www.fluigent.com/resources-support/support-tools/microfluidic-calculators/shear-stress-calculator/}.

\bibitem{fois_dynamic_2021}
Fois CAM, Schindeler A, Valtchev P, Dehghani F.
\newblock Dynamic flow and shear stress as key parameters for intestinal cells morphology and polarization in an organ-on-a-chip model.
\newblock Biomed Microdevices. 2021;23(4):55.
\newblock doi:{10.1007/s10544-021-00591-y}.

\bibitem{shao_integrated_2009}
Shao J, Wu L, Wu J, Zheng Y, Zhao H, Jin Q, et~al.
\newblock Integrated microfluidic chip for endothelial cells culture and analysis exposed to a pulsatile and oscillatory shear stress.
\newblock Lab Chip. 2009;9(21):3118.
\newblock doi:{10.1039/b909312e}.

\bibitem{kim_dynamic_2024}
Kim S, Lam PY, Richardson LS, Menon R, Han A.
\newblock A dynamic flow fetal membrane organ-on-a-chip system for modeling the effects of amniotic fluid motion.
\newblock Biomed Microdevices. 2024;26(3):32.
\newblock doi:{10.1007/s10544-024-00714-1}.

\bibitem{Poon}
Poon C.
\newblock Measuring the density and viscosity of culture media for optimized computational fluid dynamics analysis of in vitro devices.
\newblock bioRxiv. 2020;doi:{10.1101/2020.08.25.266221}.

\bibitem{corral-najera_polymeric_2023}
Corral-Nájera K, Chauhan G, Serna-Saldívar SO, Martínez-Chapa SO, Aeinehvand MM.
\newblock Polymeric and biological membranes for organ-on-a-chip devices.
\newblock Microsyst Nanoeng. 2023;9(1):107.
\newblock doi:{10.1038/s41378-023-00579-z}.

\bibitem{strods_development_2024}
Strods A, Narbute K, Movčana V, Gillois K, Rimša R, Hollos P, et~al.
\newblock Development of {Organ}-on-a-{Chip} {System} with {Continuous} {Flow} in {Simulated} {Microgravity}.
\newblock Micromachines. 2024;15(3):370.
\newblock doi:{10.3390/mi15030370}.

\bibitem{wang_new_2023}
Wang L, Chen Z, Xu Z, Yang Y, Wang Y, Zhu J, et~al.
\newblock A new approach of using organ-on-a-chip and fluid–structure interaction modeling to investigate biomechanical characteristics in tissue-engineered blood vessels.
\newblock Front Physiol. 2023;14:1210826.
\newblock doi:{10.3389/fphys.2023.1210826}.

\bibitem{Stokes1851}
Stokes GG.
\newblock On the Effect of the Internal Friction of Fluids on the Motion of Pendulums.
\newblock Transactions of the Cambridge Philosophical Society. 1851;9:8.

\bibitem{batchelor1967introduction}
Batchelor GK.
\newblock An Introduction to Fluid Dynamics.
\newblock Cambridge University Press; 1967.
\newblock Available from: \url{https://www.cambridge.org/core/books/an-introduction-to-fluid-dynamics/18AA1576B9C579CE25621E80F9266993}.

\bibitem{Ostwald1902}
van Laar JJ.
\newblock Ober einen Aufsatz des Herrn Schükarew.
\newblock Zeitschrift für Physikalische Chemie. 1902;39U(1):342--344.
\newblock doi:{doi:10.1515/zpch-1902-3923}.

\bibitem{Griffiths2020}
Griffiths PT.
\newblock {Non-Newtonian channel flow—exact solutions}.
\newblock IMA Journal of Applied Mathematics. 2020;85(2):263--279.
\newblock doi:{10.1093/imamat/hxaa005}.

\bibitem{carreau1972}
Carreau PJ.
\newblock Rheological equations from molecular network theories.
\newblock Transactions of the Society of Rheology. 1972;16(1):99--127.

\bibitem{stokes1845theories}
STOKES G.
\newblock On the theories of internal friction of fluids in motion.
\newblock Trans Camb Philos Soc. 1845;8:287--305.

\bibitem{hagen1839flow}
Hagen G.
\newblock On the flow of water in narrow cylindrical tubes.
\newblock Ann Phys Chem. 1839;46:423--442.

\bibitem{poiseuille1844recherches}
Poiseuille JL.
\newblock Recherches exp{\'e}rimentales sur le mouvement des liquides dans les tubes de tr{\`e}s-petits diam{\`e}tres.
\newblock Imprimerie Royale; 1844.

\bibitem{comsol}
COMSOL. Multiphysics\textsuperscript{\textregistered} v. 6.0;.
\newblock Available from: \url{www.comsol.com}.

\bibitem{birjukovs2021resolving}
Birjukovs M, Trtik P, Kaestner A, Hovind J, Klevs M, Gawryluk DJ, et~al.
\newblock Resolving Gas Bubbles Ascending in Liquid Metal from Low-{SNR} Neutron Radiography Images.
\newblock Applied Sciences. 2021;11(20).
\newblock doi:{10.3390/app11209710}.

\bibitem{birjukovs-particle-EXIF}
Birjukovs M, Zvejnieks P, Lappan T, Sarma M, Heitkam S, Trtik P, et~al.
\newblock Particle tracking velocimetry in liquid gallium flow around a cylindrical obstacle.
\newblock Experiments in Fluids. 2022;63.
\newblock doi:{10.1007/s00348-022-03445-2}.

\bibitem{birjukovs-solidification-image-processing}
Birjukovs M, Shevchenko N, Eckert S.
\newblock An image processing pipeline for in situ dynamic x-ray imaging of directional solidification of metal alloys in thin cells.
\newblock Experiments in Fluids. 2023;64.
\newblock doi:{10.1007/s00348-023-03671-2}.

\bibitem{local-adaptive-thresholding}
Sezgin M, Sankur B.
\newblock Comparison of thresholding methods for non-destructive testing applications.
\newblock Journal of Electronic Imaging. 2004;13(1):46–165.

\bibitem{otsu-thresholding}
Otsu N.
\newblock A Threshold Selection Method from Gray-Level Histograms.
\newblock Systems, Man and Cybernetics, IEEE Transactions on. 1979;9:62--66.

\bibitem{images-mathematical-morphology}
Haralick R, Sternberg S, Zhuang X.
\newblock Image Analysis Using Mathematical Morphology.
\newblock Pattern Analysis and Machine Intelligence, IEEE Transactions on. 1987;PAMI-9:532 -- 550.
\newblock doi:{10.1109/TPAMI.1987.4767941}.

\bibitem{reproduction-of-color-chapter-6}
Hunt RWG.
\newblock 6.
\newblock In: Tone Reproduction. John Wiley \& Sons, Ltd; 2004. p. 47--67.
\newblock Available from: \url{https://onlinelibrary.wiley.com/doi/abs/10.1002/0470024275.ch6}.

\bibitem{wolfram-brightness-equalize}
Research W. BrightnessEqualize; 2017.
\newblock Available from: \url{https://reference.wolfram.com/language/ref/BrightnessEqualize.html}.

\bibitem{perona-malik}
Perona P, Malik J.
\newblock Scale-space and edge detection using anisotropic diffusion.
\newblock IEEE Transactions on Pattern Analysis and Machine Intelligence. 1990;12(7):629--639.
\newblock doi:{10.1109/34.56205}.

\bibitem{weickert-nonlinear-aniso-diff-schemes}
Weickert J, Romeny BMTH, Viergever MA.
\newblock Efficient and reliable schemes for nonlinear diffusion filtering.
\newblock IEEE Transactions on Image Processing. 1998;7(3):398--410.
\newblock doi:{10.1109/83.661190}.

\bibitem{kapur-entropy-segmentation}
Kapur JN, Sahoo P, Wong AKC.
\newblock A new method for gray-level picture thresholding using the entropy of the histogram.
\newblock Computer Vision, Graphics, and Image Processing. 1980;29:273--285.
\newblock doi:{10.1016/0734-189X(85)90125-2}.

\end{thebibliography}

\end{document}